\documentclass[twocolumn]{aastex62}
\usepackage{lineno}
\usepackage{soul}
\graphicspath{{./}{figures/}}

\submitjournal{ApJ}

\shorttitle{Modeling ICME with internal magnetic field}
\shortauthors{Provornikova et al.}
\begin{document}

\title{MHD modeling of a geoeffective interplanetary CME with the magnetic topology informed by in-situ observations}

\correspondingauthor{Elena Provornikova}
\email{Elena.Provornikova@jhuapl.edu}

\author[0000-0001-8875-7478]{Elena Provornikova}
\affil{The Johns Hopkins University Applied Physics Laboratory, Laurel, MD 20723, USA}

\author[0000-0003-4344-5424]{Viacheslav G. Merkin}
\affil{The Johns Hopkins University Applied Physics Laboratory, Laurel, MD 20723, USA}

\author[0000-0002-8164-5948]{Angelos Vourlidas}
\affil{The Johns Hopkins University Applied Physics Laboratory, Laurel, MD 20723, USA}

\author[0000-0001-5066-8509]{Anna Malanushenko}
\affil{High Altitude Observatory, National Center for Atmospheric Research, Boulder, CO, USA}

\author[0000-0001-9831-2640]{Sarah E. Gibson}
\affiliation{High Altitude Observatory, National Center for Atmospheric Research, Boulder, CO, USA}

\author[0000-0001-5226-2107]{Eric Winter}
\affil{The Johns Hopkins University Applied Physics Laboratory, Laurel, MD 20723, USA}

\author[0000-0001-9326-3448]{Nick Arge}
\affil{NASA Goddard Space Flight Center, Greenbelt, MD 20771, USA}

\begin{abstract}
Variations of the magnetic field within solar coronal mass ejections (CMEs) in the heliosphere depend on the CME's magnetic structure as it leaves the solar corona and its subsequent evolution through interplanetary space. To account for this evolution, we developed a new numerical model of the inner heliosphere that simulates the propagation of a CME through a realistic background solar wind and allows various CME magnetic topologies. To this end, we incorporate the Gibson-Low CME model within our global MHD model of the inner heliosphere, GAMERA-Helio. We apply the model to study the propagation of the geoeffective CME that erupted on 3 April, 2010 with the aim to reproduce the temporal variations of the magnetic field vector during the CME passage by Earth. Parameters of the Gibson-Low CME are informed by STEREO white-light observations near the Sun. The magnetic topology for this CME—the tethered flux rope—is informed by in-situ magnetic field observations near Earth. We performed two simulations testing different CME propagation directions. For an in-ecliptic direction, the simulation shows a rotation of all three magnetic field components within the CME, as seen at Earth, similar to that observed. With a southward propagation direction, suggested by coronal imaging observations, the modeled $B_y$ and $B_z$ components are consistent with the {\it ACE} data, but the $B_x$ component lacks the observed change from negative to positive. In both cases, the model favors the East-West orientation of the CME flux rope, consistent with the orientation previously inferred from the {\it STEREO/HI} heliospheric images.
\end{abstract}

\keywords{Sun  --- 
coronal mass ejection --- magnetohydrodynamics --- shock waves}

%
%
%
%
%INTRODUCTION
%
%
%
%
\section{Introduction} \label{sec:intro}
Observations indicate that the internal magnetic structure of coronal mass ejections (CMEs) released from the Sun into the heliosphere can be well described as a magnetic flux rope \citep{vourlidas2013many}. 
Even more convincing evidence of the flux rope structure in CMEs has recently emerged through detailed close-up imaging observations of multiple CMEs by the WISPR coronograph on Parker Solar Probe \citep{howard2022overview}. 
As the CME propagates from the Sun, its large-scale twisted magnetic field introduces a large disturbance in the interplanetary magnetic field. %that in quiet times can be well-described as the  radial field with a corotating azimuthal component.
When they reach Earth, CMEs can generate geomagnetic storms with the storm intensity regulated in part by the magnetic field within the CME, in particular, the  magnitude and duration of the southward $B_z$ component.  Physical understanding of the magnetic structure of CMEs, including their formation on the Sun and evolution in the heliosphere, and the need to forecast magnetic field variations during the CME passage by Earth has long been recognized as an important focus of research.

Global magnetohydrodynamic (MHD) simulations have long been used as a tool to understand the physics of the CME evolution in the corona and the inner heliosphere, and to predict their space weather-relevant characteristics, e.g., the time-of-arrival  at the Earth. 
Physics-based Sun-to-Earth simulations start from the solar surface and model the CME in the corona and further propagation to 1~au \citep{torok2018sun, Merkin2016b, manchester:2008, lugaz2005evolution}. \citet{torok2018sun} simulated the extreme 2000 July 14 ``Bastille Day" CME eruption from its pre-eruptive state in the solar corona to its passage by Earth. The physics of this simulation included a careful treatment of the initially stable magnetic configuration in the thermodynamic solar corona, from which the eruption originated, and the processes that account for heating and cooling of the solar wind and the CME plasma. The coronal model was coupled with a more physically simple heliospheric model to propagate the CME to 1~au. In this approach, the
CME that propagated through the heliosphere was produced self-consistently from the initial pre-eruptive magnetic configuration and the dynamic evolution in the corona. 
Computational resources required for such simulations remain substantial, making it currently impractical for routine runs for operational space weather forecasts.  

Another type of global MHD simulations focuses on CME propagation in the inner heliosphere. Such simulations start at a certain distance above the Sun, typically at $\sim 20\, R_{S}$ ($\sim 0.1 \,au$) beyond the Alfv\'en surface in the solar wind. The solar wind conditions are driven by the output from the Wang-Sheeley-Arge (WSA) model, combining a potential magnetic field solution with an empirical solar wind speed specification~\citep{arge:2004, arge:2000, mcgregor:2011, henney:2012}. In this case,  heliospheric models make simplifying assumptions about the initial structure of the CME. While they neglect the complex and important physics of CME eruption and evolution in the corona, they allow the quantification of effects related to the evolution of the interplanetary CME in the heliosphere due to its interaction with the background solar wind and magnetic field. For a moderate-resolution computational grid that is sufficient for resolving large-scale CMEs and solar wind streams, these models require significantly less computational resources than physics-based Sun-to-Earth simulations, making them more practical for space weather needs.

Different types of CME models have been included in such simulations of the inner heliosphere.
A simplified hydrodynamic description of a CME as a perturbation in plasma density and speed has been used to estimate CME time-of-arrival at Earth and variations of density and velocity \citep{mays2015ensemble, odstrcil2004numerical}. This approach neglects the internal CME magnetic field. Currently, the NOAA Space Weather Prediction Center uses the WSA-ENLIL heliospheric model with a hydrodynamic CME Cone model for CME forecasting purposes \citep{wold2018verification}. Observational evidence of magnetic flux ropes in CMEs, the importance of internal magnetic field  structure for the CME evolution in the heliosphere, and the need for space weather predictions of the CME magnetic field at Earth has motivated the development of models that include CMEs with an  internal  magnetic field in the heliospheric domain. Previous models in this category included a magnetic spheromak \citep{verbeke2019evolution, singh2020application, palmerio2023modeling} or a croissant-like volume with a magnetic flux rope \citep{maharana2022implementation, singh2022ensemble}. 

In this paper, we expand on the previous work by including in our global model of the inner heliosphere a CME model that allows various types of magnetic topologies. Specifically, the CME is described by the analytical Gibson-Low model \citep[hereinafter referred to as GL]{gibson:1998}, which allows topologies ranging from detached magnetic spheromak to anchored magnetic arcade -- resulting from an inward radial stretching of a basic spheromak solution. The distinct topologies produce different behavior of magnetic field components within the ejecta. This approach provides more flexibility in finding  magnetic structures matching observed CMEs, allowing for more complex topologies than a spheromak or a croissant-type flux rope. Therefore, our objective in this work is to investigate whether the magnetic field variations observed during the CME passage at Earth can be reproduced by a global MHD simulation with an appropriate choice of the GL magnetic topology.

Our global inner heliosphere model uses the Grid Agnostic MHD with Extended Research Applications (GAMERA) code with the ability to emerge a GL magnetic flux rope through its inner boundary above the corona. This is the first demonstration of a GAMERA simulation of an interplanetary CME following previous efforts focused on steady-state solar wind simulations in the inner heliosphere \citep{mostafavi2022high, Knizhnik2024}. 
The GAMERA code offers several algorithmic advances compared to other MHD codes~\citep{zhang:2019}. High resolving power, meaning that a minimal number of cells is needed to resolve a structure of a particular size, leads to relatively modest computational resources required by GAMERA to produce simulations of sufficiently high resolution. The use of constrained transport \citep{evans1988simulation}  intrinsically maintains zero divergence of the magnetic field throughout the simulation domain and at all time, including while the CME is emerging. These aspects play a key role in the development of a computationally efficient model of a magnetized CME, resolving its interaction with the background solar wind and retaining its structure during its propagation through the interplanetary space.

In this paper, we use the model to simulate the interplanetary propagation of a geoeffective CME  that erupted on 3 April, 2010. At Earth, this CME was associated with a rotation of all three magnetic field components in the ejecta and a long-lasting period of negative $B_z$ component, which we attempt to reproduce in this work. In Section \ref{sec:CME} we describe observations of the 3 April, 2010 CME event that inform the initial GL CME configuration in our model. We briefly discuss previous simulation efforts with a hydrodynamic description of a CME. Section \ref{sec:model} introduces the GAMERA model of the inner heliosphere, provides a brief description of the Gibson-Low CME model, and outlines the methodology of GAMERA and GL model coupling. Section \ref{sec:results} presents the results of the GAMERA simulations and compares the modeled temporal variations of CME magnetic field for two different propagation directions to the in-situ data. We discuss our results in Section \ref{sec:Discussion} and conclude this work in Section \ref{sec:Summary}.

%
%
%
%
%OBSERVATIONS
%
%
%
%
\section{3 April 2010 CME Event} \label{sec:CME}

\begin{figure*}[ht]
%\plotone{fig_wl_obs.png}
\centering 
\includegraphics[scale=0.9]{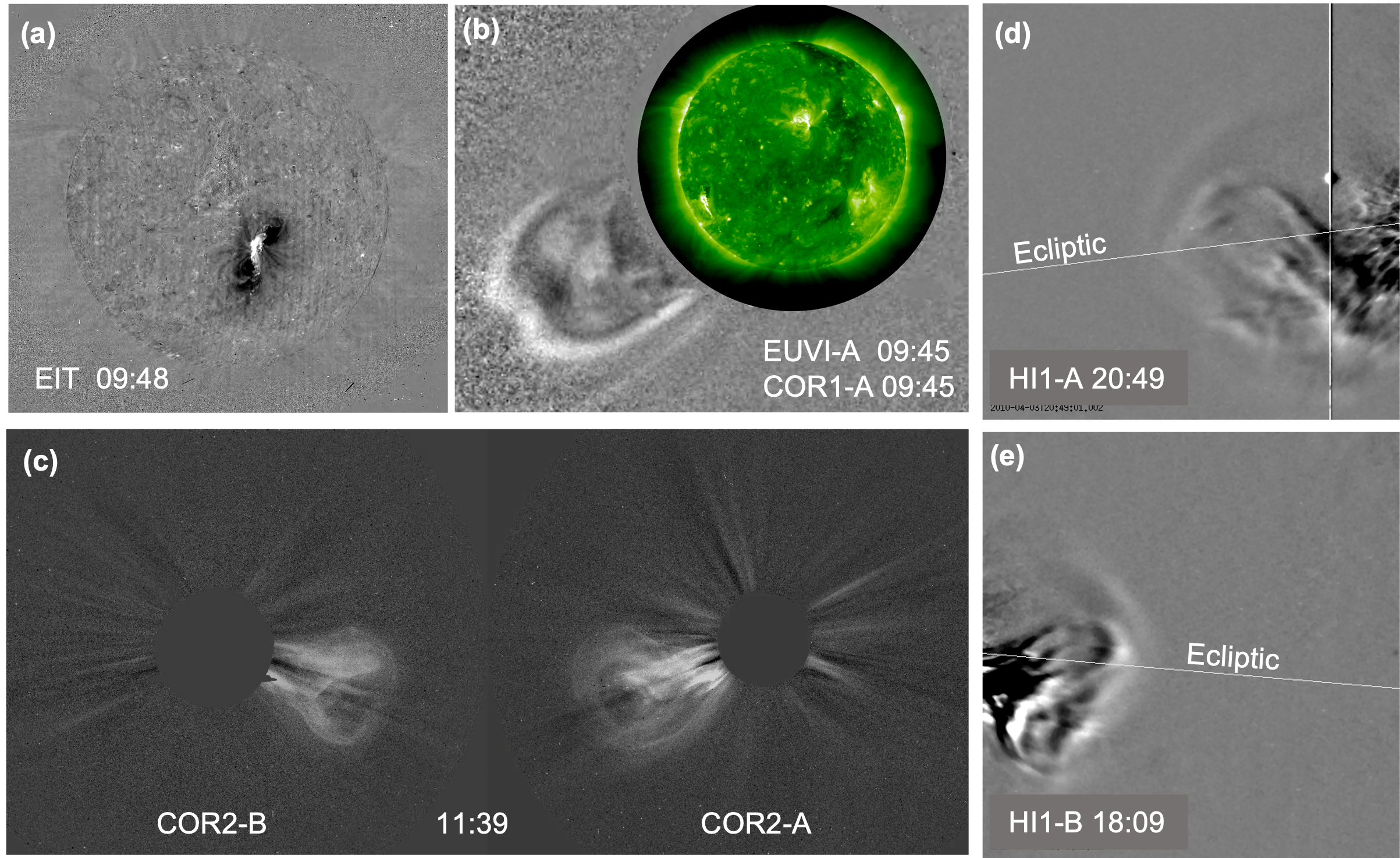}
\caption{Summary of key imaging observations of the April 3, 2010 CME. (a): SOHO/EIT 195{\AA} channel running difference image showing the source region and the extent of the post-CME arcade; (b) Composite of STA/EUVI-195\AA\ and COR1 running difference images showing the shape and direction of the CME in the inner corona; (c) Base difference COR2-B and COR2-A snapshots showing the CME expanding northward towards the ecliptic in the middle corona; (d-e) HI1-A and HI1-B running difference snapshots showing that the northern part of the CME has crossed the ecliptic plan and the formation of a wave ahead of the CME.}
\label{fig:WL_obs}
\end{figure*}

The CME on 3 April 2010 was the first geoeffective CME of Cycle 24 and received considerable attention in the literature \citep{wood2011, rouillard2011interpreting, deforest20173d,braga2017pseudo}. It was observed by coronagraphs and imagers from three viewpoints, STEREO-A (STA), -B (STB), and SOHO/LASCO. STA and STB were at near-quadrature relative to Earth, which resulted in multi-point observations for this Earth-directed CME and some ambiguity regarding its orientation, as we discuss below. This event has been described in detail in the papers mentioned above. We provide only a short description here for completeness. The CME originated in AR~11059, located at S23W10 (Figure~\ref{fig:WL_obs}a). The SECCHI/COR1 and COR2 images indicate a three-part CME, consistent with an eruption of a magnetic flux rope (MFR, Figure~\ref{fig:WL_obs}b,c). The COR2 MultiView Catalog \citep{Vourlidas_etal_2017} reports a width of $42^{\circ}-46^{\circ}$ and a speed of about 800 km/s at $\sim 10 R_S$ from the Sun center. The SECCHI/HI images (Figure~\ref{fig:WL_obs}d,e) show the development of a well-defined compression front ahead of the CME MFR \citep{rouillard2011interpreting}. 

\citet{wood2011} reported a 3D reconstruction of the CME using white-light images from all three spacecraft and found that the CME orientation could not be established unambiguously.
The SECCHI and LASCO white-light images could be modeled reasonably well by either North-South (N-S; $\phi = -80^{\circ}$, where $\phi$ is the MFR orientation angle to the ecliptic plane) or East-West (E-W; $\phi = 10^{\circ}$) orientations of the CME MFR. The trajectory of the flux rope was determined to be directed $2^{\circ}$W of Earth and $16^{\circ}$S. Based on the two orientations considered by \citet{wood2011}, Earth was hit either by the northern leg of the N-S oriented MFR or by the top of the apex of the E-W oriented MFR.

The CME with its shock wave was detected in situ by the {\it Wind} and {\it ACE} spacecraft on 5 April 2010 but the in situ measurements could not help discriminate between the two orientations because there were no unambiguous signatures of an MFR \citep{mostl2010stereo}. Although the CME showed signatures of a magnetic cloud (MC) \citep{burlaga1981magnetic}, no clear helical magnetic field structure was identified such as those that usually characterize the MC \citep{mostl2010stereo}. The CME was described as having two regions: (1) a region where the magnetic field $\vec B$ rotates and its strength $|B|$ declines, and (2) a region where $\vec B$ does not rotate and $|B|$ is constant. The second region is associated with an extended period of negative $B_z$ that drove a geomagnetic storm at Earth. \citet{mostl2010stereo} analyzed this magnetic field behavior and
concluded that {\it Wind} sampled a part of the CME where the field was not necessarily helical.   
%The Earth encounter was a 'flank' hit consistent with the location of the source region south of the ecliptic and the 3D reconstruction results. 
%It was not possible to infer one unique orientation for this CME. 
 %and then decelerates to $\sim 700$ km/s near Earth \citep{wood2011}. 
%Imaging data suggest that the CME stays connected to the Sun as it propagates toward the Earth. 
% this supports our modeling approach because we use tethered flux rope always connected to the origin.
%significant uncertainties in the orientation of many flux ropes observed on STEREO (Thernisien et al. 2009)
%Wood pointed out that self-similar expansion in their model work relatively well for this CME.

Previous simulations of this CME  focused mainly on understanding the evolution of the sheath region and the CME shock properties in the inner heliosphere. Using an MHD model developed by \citet{wu2007three}, \citet{wood2011} initiated a CME as a velocity pulse with duration and magnitude to match the arrival time of the CME front at Earth. The model showed flattening of the CME front in the southern part, which is in agreement with STEREO HI images. 
\citet{rouillard2011interpreting} studied the evolution of the CME-driven shock from 10 $R_S$ to 1 au and its relation with observed properties of solar energetic particle event associated with this CME. They modeled the solar wind structure from 10 Rs to 1 au using the ENLIL model \citep{odstrvcil1999three}. The CME was simulated with as a hydrodynamic structure in which a sudden plasma pressure increase was applied to a spherically shaped volume. As a result, a compression region with a shock formed ahead of the driver gas. 
%Their simulation was focused on understanding the spatial and temporal characteristics of the CME-driven shock.
%such as orientation relative to the background magnetic field, compression ratio, shock speed, and magnetic connectivity of the shock to ACE, Wind, STA, and STB spacecraft. 
Using this simulation, \citet{rouillard2011interpreting} obtained a relatively good agreement in the angle between the ambient heliospheric magnetic field and the shock normal and in the shock compression ratio between the ENLIL model and in-situ observations at L1. 

These simulation efforts have provided important insights into the spatial extent of the CME sheath and the properties of the driven shock. However, to understand the CME magnetic structure and its evolution in the heliosphere, it is necessary to use a modeling approach that accounts for the internal CME magnetic field. By doing so, we can determine, for instance, which part of the CME's magnetic structure was encountered by the near-Earth spacecraft and caused an extended period of negative $B_z$.

In this study, we use such approach and utilize both white-light imaging observations of a CME near the Sun and in-situ magnetic field measurements during the CME passage at 1 au to establish constraints on the parameters governing the magnetic topology, speed, orientation, and direction of the CME structure in our model. The 3 April, 2010 CME event is an attractive case for this investigation, as it was well-observed and unaffected by interactions with other CMEs.
%
%
%
%
%
%
%SECTION 3 MODELS
%
%
%
%
%
%
\section{A model of the inner heliosphere including an ICME with internal magnetic structure} \label{sec:model}

We model the propagation of the 3 April, 2010 CME in the inner heliosphere from $21.5\, R_S$ ( $0.1 \, au$) to $\approx 215\, R_S$ ($1 \, au$),
using the heliospheric adaptation of the GAMERA MHD code coupled with the GL CME model. 
In this section, we first describe the GL model (\ref{subsec:glmodel}) and discuss how we constrain the GL input parameters to model the specific CME event. Then, in Section \ref{subsec:gamhelio} we describe the GAMERA-Helio model of the inner heliosphere. 
Finally, the coupling of GAMERA and the GL CME model is described in Section \ref{subsec:glbc}.

\subsection{The Gibson-Low CME model}\label{subsec:glmodel}
The GL CME model \citep{gibson:1998} provides an analytical three-dimensional MHD solution for the time-dependent expansion of a CME in the solar corona. The model represents the CME as a closed magnetic structure with a winding magnetic field (a classical spheromak in spherical volume, or a tear-drop-shaped stretched spheromak) that expands and pushes through the ambient open magnetic field. \citet{gibson:1998} showed that the model reproduces the CME appearance in white-light coronograph images, where CMEs typically exhibit a three-part structure: a bright leading front, a dark cavity, and a high-density structure in the cavity \citep{illing1986disruption}. Consistency with imaging observations, the inclusion of a twisted magnetic field, and the flexibility of setting various magnetic topologies make the GL CME model well-suited for studies of interplanetary CME propagation.

Because the GL model analytically describes the CME's self-similar expansion from the Sun outward, it is advantageous to integrate it into an inner heliosphere model. The GL model provides a time-dependent solution for all  MHD variables, including density, velocity, magnetic field, and thermal pressure at any distance from the Sun. Therefore, it can provide the necessary boundary conditions for CME emergence at the inner boundary of heliospheric models. 

The GL model has several input parameters that control CME properties, including the speed at the CME front, the magnetic topology, the orientation of a detached or tethered spheromak, and the angular width. The flexibility in setting these parameters enables modeling CMEs with different properties, e.g., fast or slow CMEs, strongly or weakly magnetized CMEs, and spheromak-like or flux rope CMEs. Modest computational resources required to run the coupled GAMERA and GL models at sufficient resolution make running ensembles of simulations with different input CME parameters relatively inexpensive.

\begin{table*}
\centering
\caption{Gibson-Low CME model parameters and values \label{tab:1}}
\begin{tabular}{p{4cm} p{3.5cm} p{5cm}}
\hline\hline
G\&L Parameter & Value & Justification \\
\hline
Angular width & $45^{\circ}$ & {Determined from STEREO A/B white light imaging} \\
\hline
Magnetic topology, $\Upsilon$ & 2.5 & {Uniquely determined from in-situ temporal variations of $\vec B$-components near Earth} \\
\hline
CME radial velocity & 300 km/s & {GL radial speed is added to SW background speed producing the 800 km/s CME speed observed in white light} \\
\hline
Maximum magnetic field magnitude & 100 nT & Consistent with magnetic field magnitude observed during the CME passage near Earth. \\
\hline
Orientation & $0^{\circ}$ & In-ecliptic E-W orientation (in the GL model the core axis of symmetry is directed North-South) \\
\hline
Direction (lat, lon) & Case 1: ($0^{\circ}$, $60^{\circ}$) & Case 1: In-ecliptic direction \\ & Case 2: ($-15^{\circ}$, $60^{\circ}$) & Case 2: CME direction inferred from coronal images. \\
\hline
\end{tabular}
\end{table*} 

\begin{figure*}
\centering
\includegraphics[scale = 0.5, angle=-90]{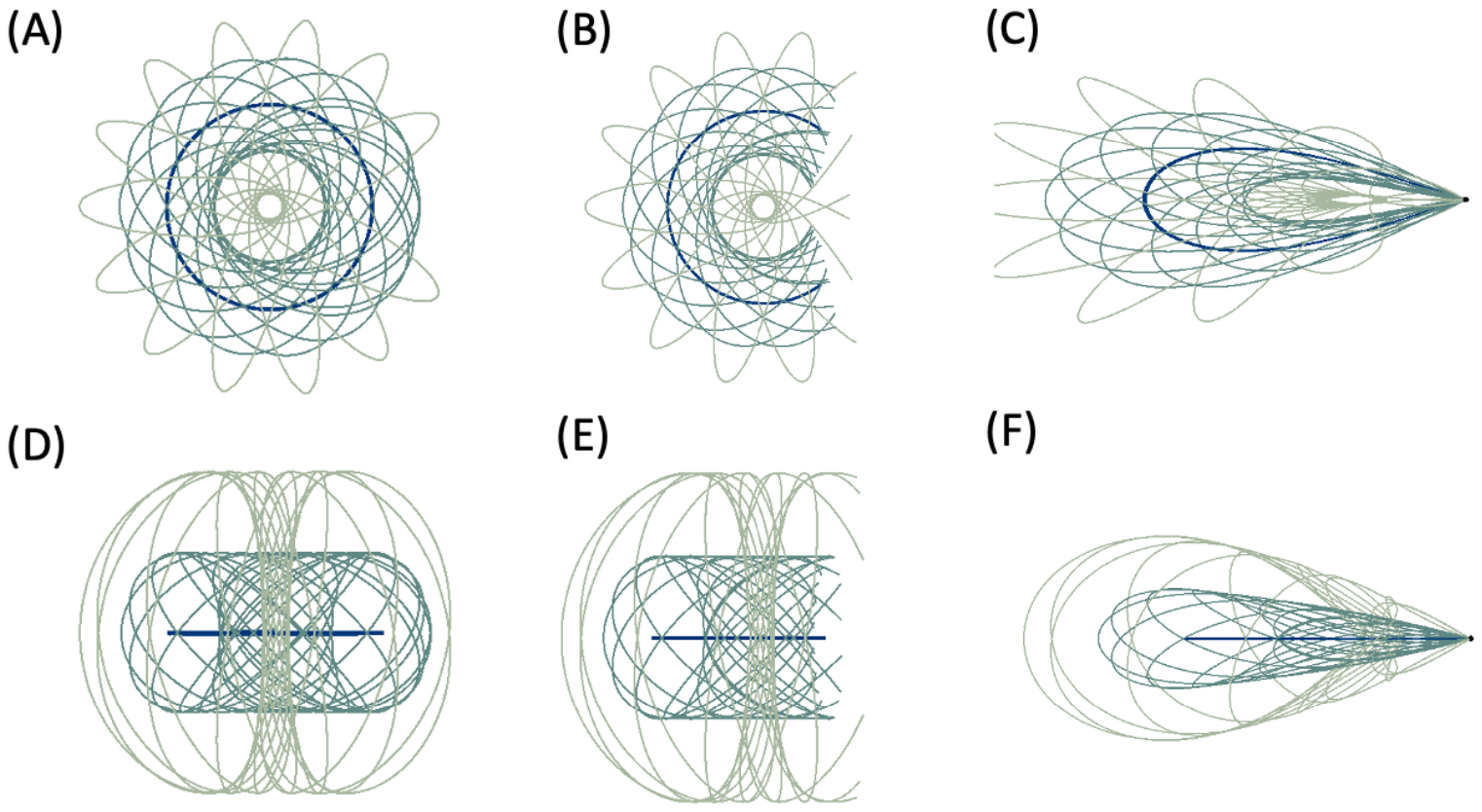}
\vspace{-0.7in}
\caption{ Transformation of a spheromak resulting in a tethered flux rope CME configuration used to simulate the 3 April, 2010 event. Panels (A), (B), (C) show the view with the axis of symmetry in the out-of-plane direction. Panels (D), (E), (F) show the view with the axis of symmetry in the plane and pointing up. Panels (A) and (D) show the initial spheromak. Panels (B) and (E) show the part of the spheromak that is transformed into a tethered flux rope after applying a radial stretching transformation. Panels (C) and (F) show the resulting tethered flux rope configuration (topmorph parameter $\Upsilon = 2.5$)}
\label{fig:tethFR}
\end{figure*}

Depending on the stretching parameter $a$ in the GL model \citep{gibson:1998} in combination with other geometric and magnetic parameters, it can produce CME solutions with different magnetic topologies, from a magnetic spheromak to a magnetic arcade, and different 3D shapes, from a spherical bubble to a tear-drop shape. For our purpose of simulating the 3 April, 2010 CME event, we have to choose the appropriate CME magnetic topology.

In recent work, \citet{malanushenko2020convolutional} introduced a new dimensionless parameter called ``topmorph'' $\Upsilon$, which defines the type of the magnetic topology and morphology of a GL CME. We here use an expression for the ``topmorph'' $\Upsilon = \frac{x_0-a}{r_0/2}+2$ \footnote{Note, this expression defines the topology relative to the center of the Sun, which differs slightly from the definition in \citet{malanushenko2020convolutional} who defined it relative to the Sun's photosphere at $r=R_S$.} where $x_0$ is the distance from the center of the Sun to the center of the CME sphere, $r_0$ is the radius of the sphere, and $a$ is the stretching parameter, using notation from Fig.~5 in \citet{gibson:1998}.  The ``topmorph'' parameter takes values $\Upsilon>0$. The value corresponds to one of the possible topologies: $\Upsilon  \geqslant 4$ is a detached (from the center of the Sun) spheromak, $3 \leqslant \Upsilon < 4$ is a tethered spheromak, $2 \leqslant \Upsilon < 3$ is a tethered flux rope, $\Upsilon < 2$ is a simple magnetic arcade. 
The larger the stretching parameter $a$ is, the smaller the ``topmorph'' value becomes, and the stronger the CME deviates from a magnetic spheromak.  
For each topology, the GL analytical solution prescribes a unique configuration of the magnetic field inside the CME and, hence, distinct temporal variations of the magnetic field components when the CME passes through a particular point in space.

Therefore, for the 3 April 2010 CME event, the time series of the three magnetic field components measured in-situ at Earth can be used to constrain our choice of the magnetic topology in the CME model, e.g., the ``topmorph'' parameter. {\it Wind} magnetic field data in the GSE system of coordinates (Earth is the origin, X points towards the Sun, Y points to solar east, and Z is normal to the ecliptic) show a rotation of three magnetic field components \citep{mostl2010stereo}. $B_x$ and $B_z$ change the sign, while $B_y$ preserves the same negative sign. Furthermore, the $B_z$ component shows a long-duration negative period. 
Such variations of the magnetic field components are consistent with only one GL magnetic topology -- a tethered flux rope, with the ``topmorph''  value $2 \leqslant \Upsilon < 3$. Therefore, we adopt $\Upsilon = 2.5$ (Table \ref{tab:1}). The resulting tethered flux rope configuration is shown in two views in Figure \ref{fig:tethFR} (C) and (F). The top row is the view with the axis of symmetry pointing out of the plane, while the bottom row has the axis of symmetry in the plane and pointing up. Panels (A) and (D) show the original spheromak. Panels (C) and (F) show the tethered spheromak which is obtained by applying a stretching transformation that contracts a part of the original spheromak shown in (B) and (E) to the origin  (the cutout parts are contracted to the origin). In the GL solution, the tethered flux rope remains connected to the Sun during its evolution.

Other parameters of the GL model must be set to define the angular width of the CME, the radial velocity at the front of the CME, the initial height of the CME front, the factor determining the maximum magnetic field strength in the CME, and its orientation relative to the ecliptic plane. The direction of the CME propagation, i.e., the latitude and longitude where the CME emerges through the inner boundary of GAMERA, is specified in the GAMERA model (Section \ref{subsec:gamhelio}). We consider two possible directions of the CME propagation based on white-light images closer and farther from the Sun. We select $45^{\circ}$ for the angular width according to the angle between the MFR legs determined from the STEREO white-light images. We also set the CME front speed at 21.5 $R_S$ to $\sim 800\, km/s$ in accordance with these observations. The initial height of the CME front is set to be at the GAMERA inner boundary $21.5\, R_S$. The maximum magnitude of the magnetic field in the CME flux rope is taken to be $100$ nT. 
This value was chosen to make the magnetic field magnitude in the CME during the Earth passage  consistent with observations.
In the absence of an unambiguous observational determination of the orientation of the flux rope for this CME (Section \ref{sec:CME}), we experimented with different values ($0^{\circ}$, $10^{\circ}$, $45^{\circ}$, and $90^{\circ}$) in our simulations. As demonstrated in the next section, the $0^{\circ}$  orientation produces variations of the $\vec B$ components most consistent with the {\it Wind} and {\it ACE} in-situ measurements. For $0^{\circ}$ orientation, the axis of symmetry of the original spheromak before the transformation (Figs. \ref{fig:tethFR} A and D) is oriented parallel to the Z-axis (along the solar rotation axis). In other words, the flux rope has an East-West orientation. 

In summary, guided by  remote and in situ observations of the 3 April, 2010 CME event, we have defined all the necessary input parameters for the GL CME model. These parameters, along with the justification, for the specific values chosen, are summarized in Table \ref{tab:1}.
%
%
%
%
%
%
%SECTION 3.1 GAMERA model for the inner heliosphere
%
%
%
%
%
%
\subsection{GAMERA model for the inner heliosphere}\label{subsec:gamhelio}
GAMERA is an MHD code which carries on the legacy of the Lyon-Fedder-Mobarry (LFM) MHD code \citep {Lyon04}. GAMERA retains all of the core numerical algorithms of LFM, with some important improvements, but the software has been rewritten in its entirety to adapt the code to modern supercomputing architectures. Details of GAMERA's numerics, along with code performance on standard MHD test problems, were extensively described by \citet{zhang:2019}. The algorithm is characterized by high-order reconstruction (7th or 8th order) and aggressive flux limiting, ensuring low numerical dissipation. Divergence-free magnetic field is maintained to within machine precision using constrained transport \citep{evans1988simulation}. GAMERA solves the MHD equations on an arbitrary hexahedral grid that can be adapted to the specific problem at hand. GAMERA  has been extensively used for global simulations of the Earth's magnetosphere \citep{michael2021modeling, sorathia2020ballooning, sorathia2021role}. \citet{mostafavi2022high} used the heliospheric application of GAMERA to model the solar wind in the inner heliosphere with a very high resolution to study the generation of the Kelvin-Helmholtz instability at flow shears. In recent work, \citet{Knizhnik2024} performed solar wind simulations in the inner heliosphere using the GAMERA model with boundary conditions driven by two different photospheric flux transport models and compared the GAMERA results with multi-point in-situ solar wind measurements. The work presented here describes a new capability of the heliospheric version of GAMERA to simulate the propagation of CMEs in the inner heliosphere. 

The GAMERA setup for  simulation of the background solar wind follows the previous approach used with the heliospheric version of LFM, LFM-Helio~\citep{merkin_hcs_fold:2011, merkin:2016}. In this paper, GAMERA solves the ideal MHD equations on a spherical grid in the inertial Cartesian Sun-centered system of coordinates. The coordinate system is defined such that the X-axis is aligned with the Sun-Earth line pointing away from Earth such that at $t=0$ in the simulation Earth is located at azimuthal angle $\phi = 180^{\circ}$, Z-axis coincides with the solar rotation axis pointing to the solar north and the Y-axis completes the right-handed coordinate system.  The grid extends from the inner boundary at $r=21.5\,R_S$ to the outer boundary at Earth's orbit, $r=215\,R_S$, $18^{\circ}  \leqslant \theta \leqslant 162^{\circ}$ in the latitudinal direction with $\theta = 0^{\circ}$ corresponding to solar north, and $0 \leqslant \phi \leqslant 360$ in the longitudinal direction.
The grid spatial resolution used in this work is uniform ($256 \times 128 \times 256$) in $(r,\theta, \phi)$ directions, respectively.  Due to GAMERA's high resolving power, this relatively modest resolution allows resolving not only large-scale structures in the solar wind but also details of stream interaction regions, the internal magnetic structure of CMEs, CME-driven shocks, and CME sheaths, as will be demonstrated by the simulation results in Section \ref{sec:results}. 

To model the three-dimensional structure of the solar wind in the inner heliosphere, we must specify the inner boundary conditions at $r = 21.5 R_{S}$. As in previous studies with the LFM-Helio model, we use the semi-empirical WSA-ADAPT model of the solar corona \citep{arge:2000, arge:2004, mcgregor:2011, wallace2020relationship} 
which utilizes global solar photospheric field maps from the ADAPT model \citep{arge:2010, arge:2011, arge:2013, hickmann:2015}.
The WSA-ADAPT model provides latitude-longitude maps of radial components of magnetic field $B_r$ and solar wind velocity $V_r$ at $r = 21.5 R_{\odot}$, which is set to be the outer boundary of the WSA-ADAPT model and the inner boundary of GAMERA. The azimuthal component of the magnetic field $B_{\phi}$ is set assuming that the magnetic field corotates with the Sun, meaning that in the corotating frame, the electric field is zero at the boundary $\mathbf{V} \times \mathbf{B} = 0$. This yields $B_{\phi} = -\Omega R_0 sin (\theta) (B_r/V_r)$ in the inertial frame, where $\Omega = 2\pi/T_{sid}$ and $T_{sid}=25.38$ days is the sidereal rotation period of the Sun (i.g. measured relative to distant stars). The polar component of the magnetic field is zero ($B_{\theta} = 0$) at the inner boundary.
The solar wind velocity in the WSA-ADAPT model is calculated using the empirical relation: 
\begin{equation}
V_r = V_{slow} + \frac{V_{fast}}{(1+f_s)^{1/4.5}} \left( 1 - 0.8 \exp^{-(d/2)^2} \right)^3
\end{equation}
where $V_{slow} = 285\, km/s$ and $V_{fast}=625\, km/s$ are the nominal speeds of the slow and fast solar wind. Parameters $f_s$ and $d$ are related to the geometry of the coronal magnetic field and represent coronal magnetic flux tube expansion factor and shortest angular distance to the nearest coronal hole boundary, respectively. Inclusion of the parameter $d$ allows for the generation of the slow solar wind in regions with a small expansion factor away from the heliospheric current sheet (hereinafter HCS), e.g., pseudostreamers, which would otherwise result in a fast wind \citep{riley:2015}. Angular velocity components are zero $V_{\phi}=0$, $V_{\theta} = 0$ at the inner boundary. Number density is calculated using an empirical fit to Helios plasma measurements \citep{mcgregor:2011, merkin_hcs_fold:2011}:
\begin{equation}\label{den}
n [cm^{-3}] = 112.64 + 9.49 \cdot 10^7/(V_r [km/s])^2
\end{equation}
Plasma temperature is calculated assuming the total pressure balance at the inner boundary. A constant for the total pressure is determined by setting the thermal pressure in the HCS where the magnetic field is zero. Since the nominal density in the HCS is obtained by eq. (\ref{den}), we only need to set the plasma temperature in the HCS, which is a free parameter of the model.

With these inner boundary conditions, GAMERA solves ideal MHD equations in the inertial frame (note, in previous studies with the LFM-Helio model, \citet{merkin:2016} used the frame rotating with the Sun, while \citet{merkin_hcs_fold:2011,Merkin2016b} used the inertial frame.). In the inertial frame, the inner boundary condition map rotates according to the solar rotation period on a static spherical grid, resulting in time-dependent boundary conditions at $21.5\,R_S$. Note that the rotation is only applied to the boundary conditions for the solar wind background. The emerging CME is assumed to propagate radially, not influenced by the rotation. The boundary conditions for emerging a CME with an internal magnetic field are described in the next section (section~\ref{subsec:glbc}). 

%
%
%
%
%
%GAMERA-GL MODEL COUPLING
%
%
%
%
%
%
%
%
\subsection{GAMERA-GL model coupling}\label{subsec:glbc}
 
We begin to emerge the CME through the GAMERA inner boundary after the background solar wind solution achieves a steady state.
We only use the GL solution internal to the CME bubble, ignoring the GL external solution for the corona; the latter is replaced by the solar wind background solution in GAMERA driven by the WSA-ADAPT. 
Since we used GAMERA version with an 8th-order spatial reconstruction in this paper, the driving inner boundary conditions are defined in four layers of ghost cells in the radial direction below the inner boundary. At every time step of the GAMERA simulation, the GL solution is calculated in a spherical shell of these four cell layers. 

A cross-section of the emerging GL CME bubble and the spherical GAMERA inner boundary represents a circle. Boundary conditions are set depending on the location of the face of the boundary grid cell - inside or outside the circle - with a changing radius that bounds the area of emergence.
For grid cells outside the circle, boundary conditions driving the ambient solar wind background are applied (see Sec. \ref{subsec:gamhelio}). For grid cells inside the circle, boundary conditions are applied so as to drive the GL CME emergence. Inside the circle, the radial velocity provided by the GL model is added to the solar wind radial velocity: 

\[ V_r (\theta, \phi, t) = V_{WSA}(\theta, \phi, t) + V_{GL} (\theta, \phi, t) \]

Note, that the plasma motion is purely radial in the GL model.
The plasma number density inside the GL CME bubble is set to be by a factor of two larger than the value in the background solar wind.
This choice was justified by a posteriori comparison with the {\it ACE} in-situ measurements of the plasma density inside the CME. The temperature inside the CME is set equal to the temperature of the background solar wind. Note that the inner boundary conditions in GAMERA imply that no information propagates inward, so the plasma flows at $21.5\, R_S$ must be faster than the fast magnetosonic speed. Therefore, we ensure that this condition is satisfied inside the GL bubble during the entire simulation by tracking a value of the fast magnetosonic Mach number $M_f$ at the boundary.

Boundary conditions for the magnetic field in the emergence area are obtained by setting tangential components of the magnetic field, $B_{\theta}$ and $B_{\phi}$, provided by the GL solution and evolving the radial magnetic field $B_{r}$ by applying appropriate electric fields at cell edges. 
If a CME magnetic field structure satisfies a condition $\nabla \cdot \vec{B} = 0$ -- as the GL solution does -- then through Faraday's law, the magnetic field stays divergence-free over the entirety of the simulation time automatically.
%\[ \partial_t B_r = \partial_{\phi} E_{\theta} - \partial_{\theta} E_{\phi} \]

If the GL magnetic structure initially enters unmagnetized space, the radial magnetic field obtained from Faraday's law at a given point on the boundary $r_0$ is by construction equivalent to the GL internal radial field:
\begin{equation}
B_r^{GL}(t, \vec{r_0}) = \int_{0}^{t} \left[ \partial_{\phi} (V_r^{GL} B_{\phi}^{GL}) +  \partial_{\theta} (V_r^{GL} B_{\theta}^{GL}) \right]dt 
\end{equation}
where $B_{\phi}^{GL}$, $B_{\theta}^{GL}$, $V_r^{GL}$ are the tangential magnetic field components internal to the GL bubble, and the plasma velocity. Since, by construction, the internal GL magnetic field vanishes on the boundary of the magnetic bubble \citep{gibson:1998}, we emerge it into an arbitrary background magnetic field by adding the background field and the GL CME bubble field:
\begin{equation}
B_r(t, \vec{r_0}) = B_r^{WSA} (\vec{r_0}) + B_r^{GL} (t, \vec{r_0}) 
\label{eq:Bfield}
\end{equation}
 
%Electric field calculation (refer to Figure) and reference to Merkin 2016. 
%The radial component in the CME $B_r^{GL} (t, \vec{r_0})$ is evolved in time by calculating at every simulation time step the tangential components of the convective electric field, $E_{\phi}$ and $E_{\theta}$, and applying them at the edges of grid cell faces at the inner boundary of GAMERA. 

To ensure robustness of the boundary conditions described above we have tested them with a wide range of GL CME input parameters and WSA-ADAPT background solutions. Those included various CME magnetic topologies, velocities, flux rope orientations, and propagation directions, extending beyond the specific CME configurations we examine in this study. However, there are some limitations associated with this boundary conditions approach that will be discussed below in Section \ref{sec:Discussion}.

%
%
%
%
%
%SIMULATION RESULTS
%
%
%
%
%
%
%
%
\section{Simulation results} \label{sec:results} 

In this section, we present GAMERA-GL simulation results. First, we demonstrate the solution for the background solar wind through which the 3 April, 2010 CME propagates. Then, we present time-dependent boundary conditions for the CME emergence. Finally, we show and discuss the evolution of the CME in the inner heliosphere. We focus on the time series of the magnetic field components at Earth for two CME propagation directions - southward and in-ecliptic (Cases 1 and 2 in Table~\ref{tab:1}) - and compare their trends to in-situ observations.

\begin{figure*}[ht]
\centering
\includegraphics[scale=0.35]{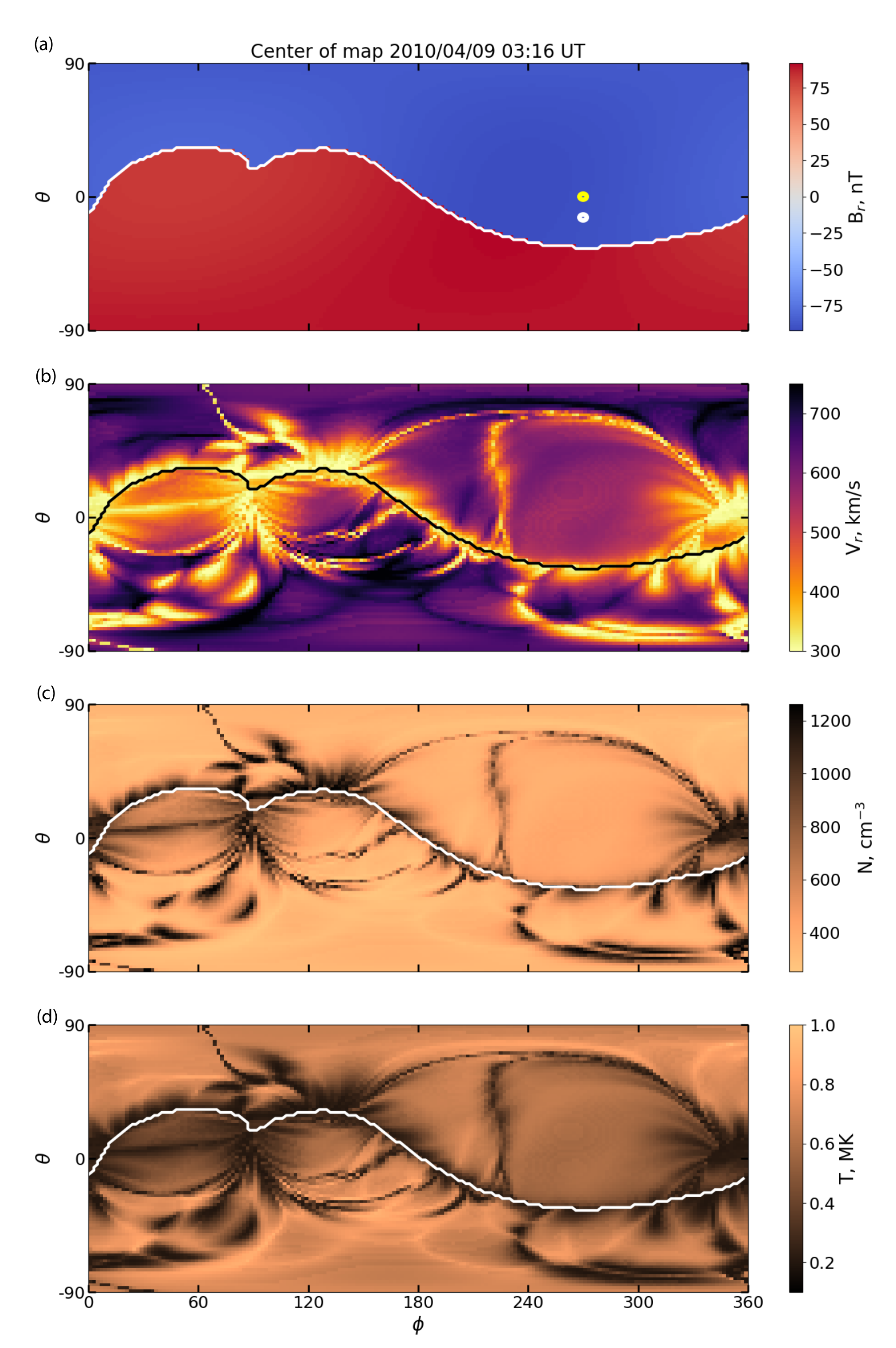}
\caption{Boundary conditions for the background solar wind at the start of the GAMERA simulation. The radial component of the magnetic field (a) and radial velocity (b), obtained from WSA-ADAPT output for CR 2095, as well as number density (c) and temperature (d), calculated as described in Section~\ref{subsec:gamhelio}, are shown as functions of latitude and longitude at a fixed radial distance, $r_{in} = 21.5 R_{S}$. The white contours in panels (a), (c) and (d) and black contour in panel (b) mark the heliospheric current sheet location defined as the $B_r=0$ iso-surface. Color scales for the density and temperature are inverted, so the faster wind is always brighter, and the slower wind is always darker. Yellow and white circles on the $B_r$ map (panel a) show in-ecliptic and southern directions of CME propagation, respectively.}
\label{fig:bc}
\end{figure*}
 
\begin{figure*}
\centering
\includegraphics[scale = 0.62, angle=-90]{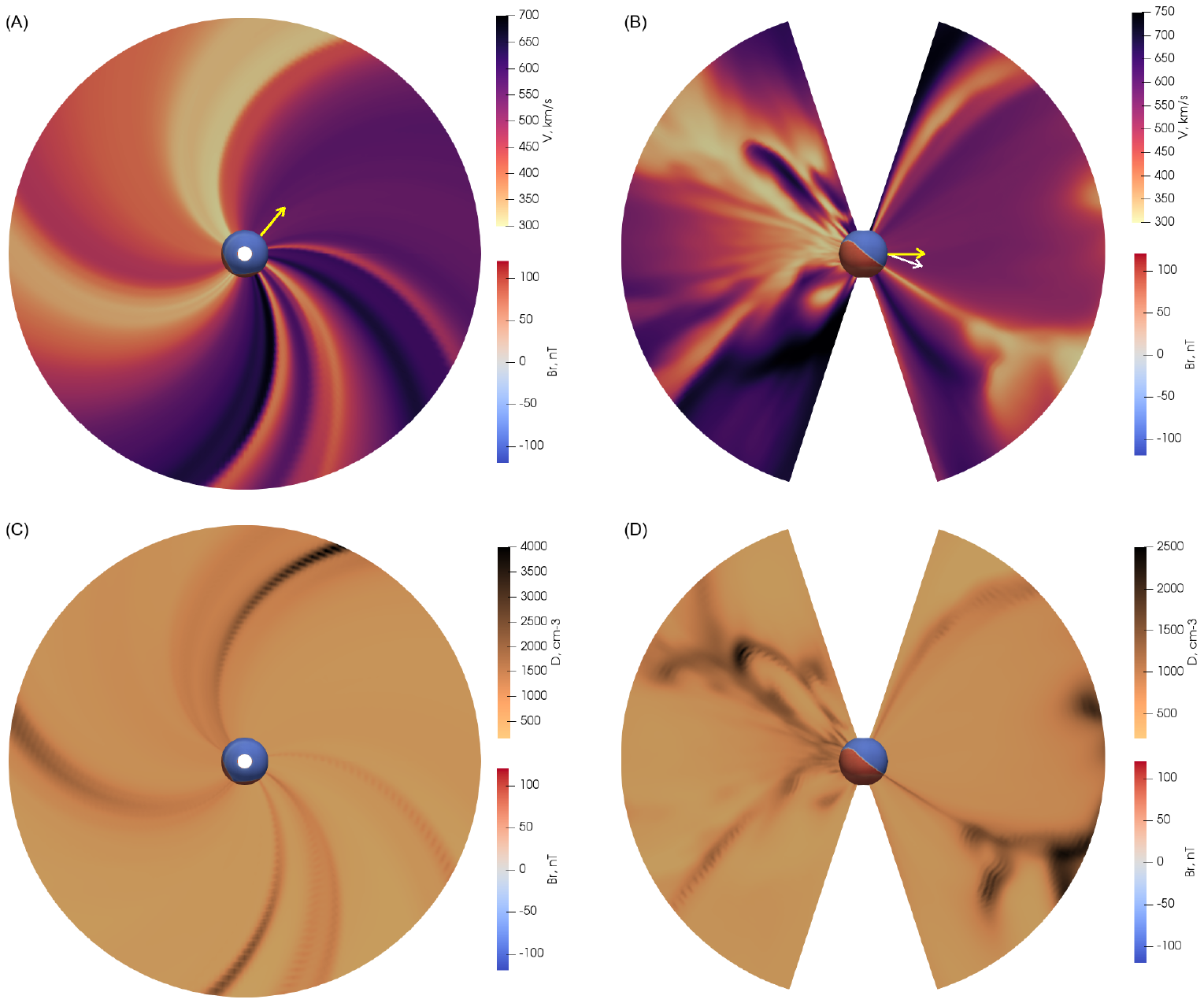}
\caption{GAMERA solution for the background solar wind in the inner heliosphere 21.5-215 $R_S$ for CR 2095. The solar wind speed $V$ and scaled solar wind density $D = n(r/r_{in})^2$  are shown in the ecliptic (A, C) and meridional (B, D) planes. The sphere in the center shows the inner boundary at 21.5 $R_S$, with the color showing $B_r$ component of the coronal magnetic field. The arrows show two propagation directions of the simulated CMEs, in-ecliptic (yellow) and $15^{\circ}$ to the south from the ecliptic (white).}
\label{fig:steadystate}
\end{figure*}

\subsection{Solar wind background}\label{subsec:solarwindbackgr}

To model the solar wind background for the 3 April, 2010 CME, we take the WSA-ADAPT output at $21.5\, R_S$ for the Carrington rotation (CR) 2095, corresponding to the 27 March--22 April, 2010 time period during the rising phase of  solar activity. The resolution of the WSA-ADAPT map is $2^{\circ} \times 2^ {\circ}$ in latitude and longitude. Figure \ref{fig:bc} shows the different variables inferred from the WSA-ADAPT maps specifying the inner boundary condition for the MHD simulation. The maps correspond to 9 April, 2010 03:16 UT, such that Earth faces the center of the map ($\phi=180^{\circ}$) at this time. The HCS ($B_r=0$ iso-surface), indicated by the white contours in panels (a,c,d) and the black contour in panel (b), is tilted relative to the solar equator. The distribution of the radial velocity shows a complex structure reflecting the conditions during the rising phase of the solar cycle. 
The slow solar wind forms a band along the HCS, with multiple arc-shaped structures away from the HCS extending to high latitudes associated with pseudostreamers.

The yellow and white circles on the $B_r$ map (panel a) mark two locations above the HCS where we emerge the CME into  the GAMERA simulation domain. 
The two latitudes represent different CME propagation directions, in-ecliptic and a southward, as suggested by imaging observations of this CME (Section \ref{sec:CME}). 

\begin{figure*}[ht]
\centering
\includegraphics[scale=0.4]{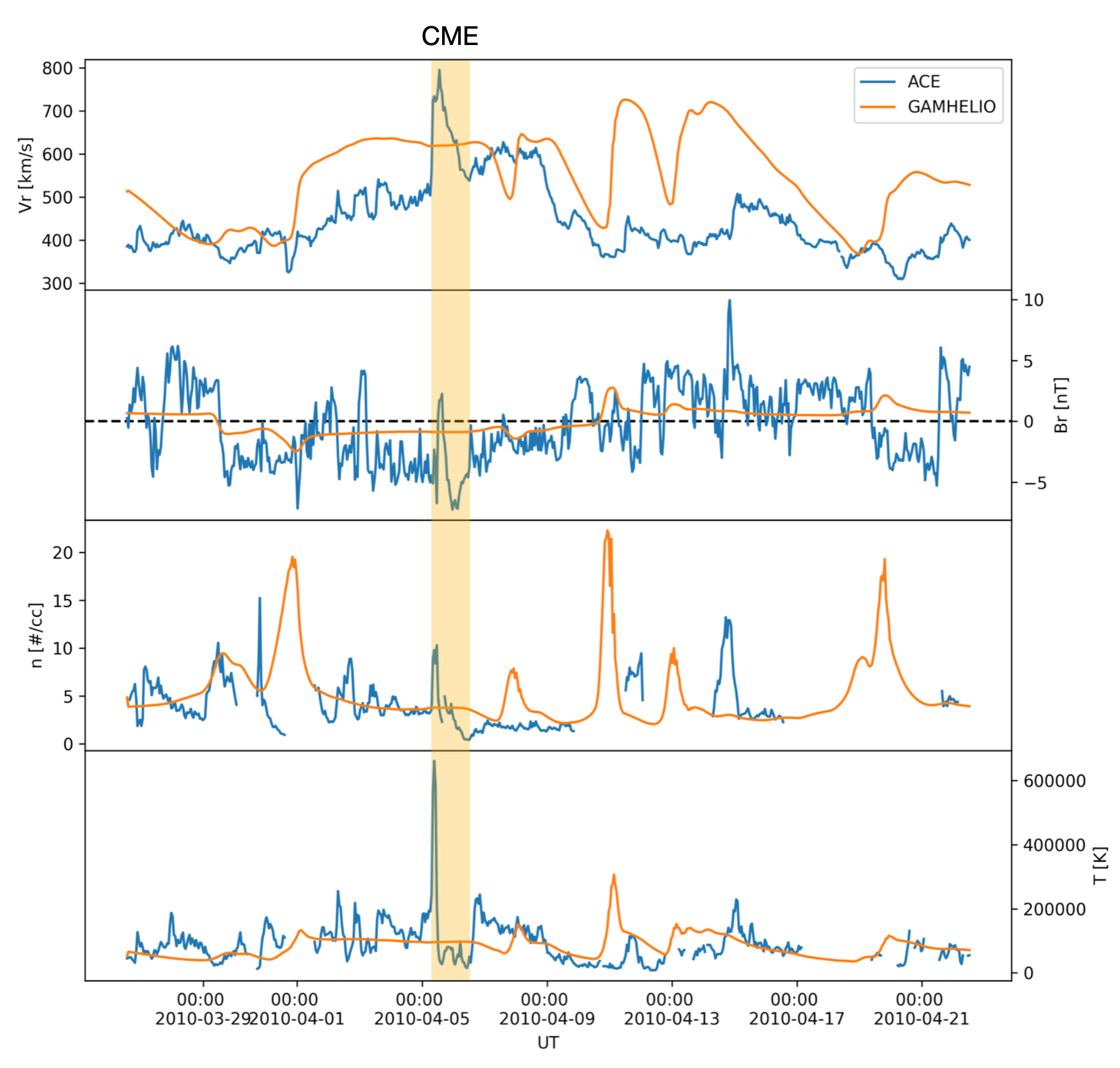}
\caption{ From top to bottom: comparison of the solar wind radial velocity, radial magnetic field component, number density, and temperature extracted along the ACE trajectory during the 27-day period in the GAMERA inner heliosphere simulation (orange curves)  with the ACE in-situ measurements (blue curves). Both the in-situ ACE data and GAMERA simulation output are taken at a 1-hr cadence. This GAMERA background solar wind simulation does not include a CME yet.}
\label{fig:ACEcomp}
\end{figure*}

Figure \ref{fig:steadystate} shows the solar wind speed and scaled density, $n(r/r_{in})^2$, in the ecliptic and meridional planar slices through the 3D MHD GAMERA solution for the inner heliosphere. For both propagation directions shown by the yellow and white arrows, CMEs with the angular width of $45^{\circ}$ will propagate in a relatively quiet solar wind region with intermediate speeds $\sim 550-650$~km/s. To see how well the model reproduces the solar wind background for the CME propagation,
Figure \ref{fig:ACEcomp} shows a comparison of the solar wind radial velocity, radial magnetic field, density, and temperature extracted from the simulation at the ACE spacecraft location and the in-situ measurements of the solar wind plasma from ACE/SWEPAM and magnetic field from ACE/MAG. 
The vertical yellow shaded region marks the CME detected by ACE on 5-7 April, 2010. It is important to note that here we are only comparing the GAMERA solar wind background simulation before we introduce the CME. Naturally, it does not yet capture the CME detected by ACE.

The CME passage at ACE shows an increase of the velocity from $\sim 500$ km/s in the preceding solar wind to $\sim 800$ km/s in the CME. Peaks in $B_r$, $n$, and $T$ in the shaded region are associated with the CME-driven shock and sheath region. A detailed analysis of in-situ measurements of this CME near Earth was reported previously by \citet{mostl2010stereo}.
In the in-situ data, the CME is embedded in the solar wind stream lasting for about 10 days (during 1-10 April, 2010) with speeds higher than the typical speed of the slow solar wind. Before the CME arrival, the solar wind speed is $\sim 500$ km/s; the speed rises to $\sim 600$ km/s after the CME passage. The GAMERA simulation  captures the large-scale solar wind stream with these intermediate speeds; however, the modeled speeds are somewhat higher before the CME arrival by $\sim 100$ km/s. In general, throughout the 27-day period, the steady-state model overestimates the solar wind speed compared to the ACE data. It is more clearly seen after 10 April, 2010, when the model shows higher speeds, meaning that in the steady-state simulation, the virtual ACE spacecraft samples the solar wind from the equatorial coronal hole, while in reality, the spacecraft had transitioned to the slow solar wind region. The later phase of the simulation is driven by the ``old'' magnetogram data, hence the lack of agreement is unsurprising. Also, 
it has been demonstrated previously by \citet{Merkin2016b}, that a time-dependent inner heliosphere model, which accounts for displacements of coronal holes, captures such transitions and shows a better agreement with the in-situ measurements than the steady-state model. 

The comparison of the modeled and measured $B_r$ of the interplanetary magnetic field shows that the model reproduces well the magnetic field polarity and captures two transitions across the HCS, from positive to negative polarities before the CME and from negative to positive polarities after the CME. Around the CME passage, the model shows the negative polarity in agreement with the ACE/MAG data. The magnitude of $B_r$ in the model is lower than in the data. The underestimation of $B_r$ component in inner heliospheric models is a well-known issue \citep{stevens2012underestimates, riley:2015, riley2019can}. This discrepancy arises from the underestimated $B_r$ component in the coronal potential field solution, which, in turn, is caused by uncertainties in the polar photospheric magnetic field measurements. The modeled density and temperature show similar values to the ACE/SWEPAM measurements, particularly during the period around the CME passage.

\begin{figure*}
\centering
\includegraphics[scale = 0.35]{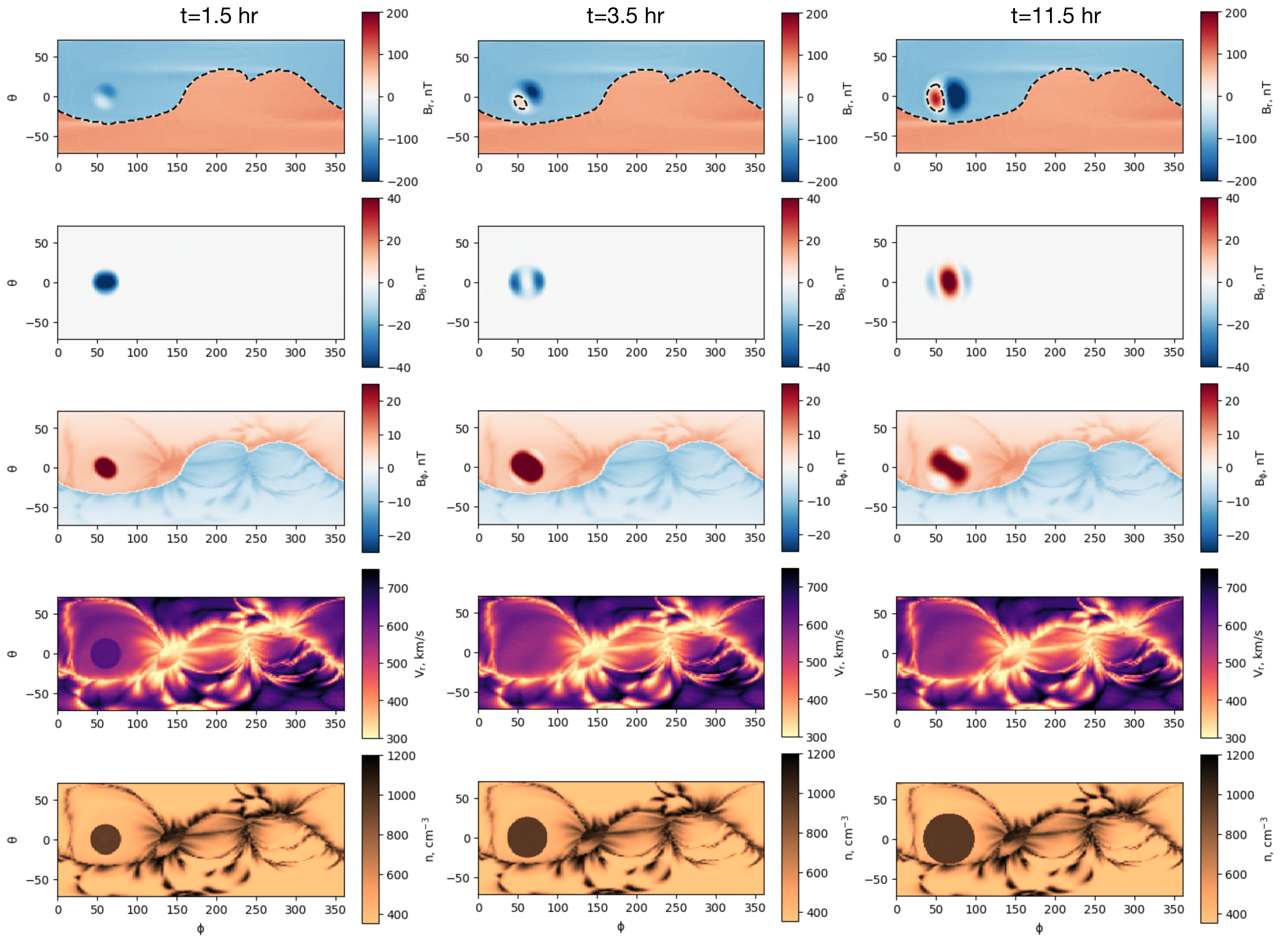}
\caption{Time-dependent GAMERA inner boundary conditions during the emergence of the Gibson-Low CME flux rope at $21.5\, R_S$. From top to bottom, panels show radial $B_r$, polar $B_{\theta},$ and azimuthal $B_{\phi}$ magnetic field components, radial velocity $V_r$ and number density $n$. Columns show increasing time in hours after the emergence started at the inner boundary $21.5\,R_S$.}
\label{fig:ibcgl}
\end{figure*}

\subsection{CME emergence and propagation}\label{subsec:cmeemergeandpropag}

Once the steady-state solution for the solar wind background is achieved, we begin to emerge the GL CME  through the inner boundary. 
Figure \ref{fig:ibcgl} shows latitude-longitude maps at $21.5\,R_S$ of magnetic field components $B_r$, $B_{\theta}$, and $B_{\phi}$, radial velocity $V_r$, and number density $n$ at different simulation times (in hours) after the CME emergence started at $21.5\,R_S$. These maps correspond to the CME propagating in the ecliptic plane (CME emergence location is centered at the yellow circle shown in Figure \ref{fig:bc}).  As noted above, the CME emerges in the solar wind of a relatively uniform speed $\sim 550-600\, km/s$ and density $\sim 500\, cm^{-3}$. The interplanetary magnetic field in the emergence area has the negative polarity. 

\begin{figure}
\includegraphics[scale = 0.2]{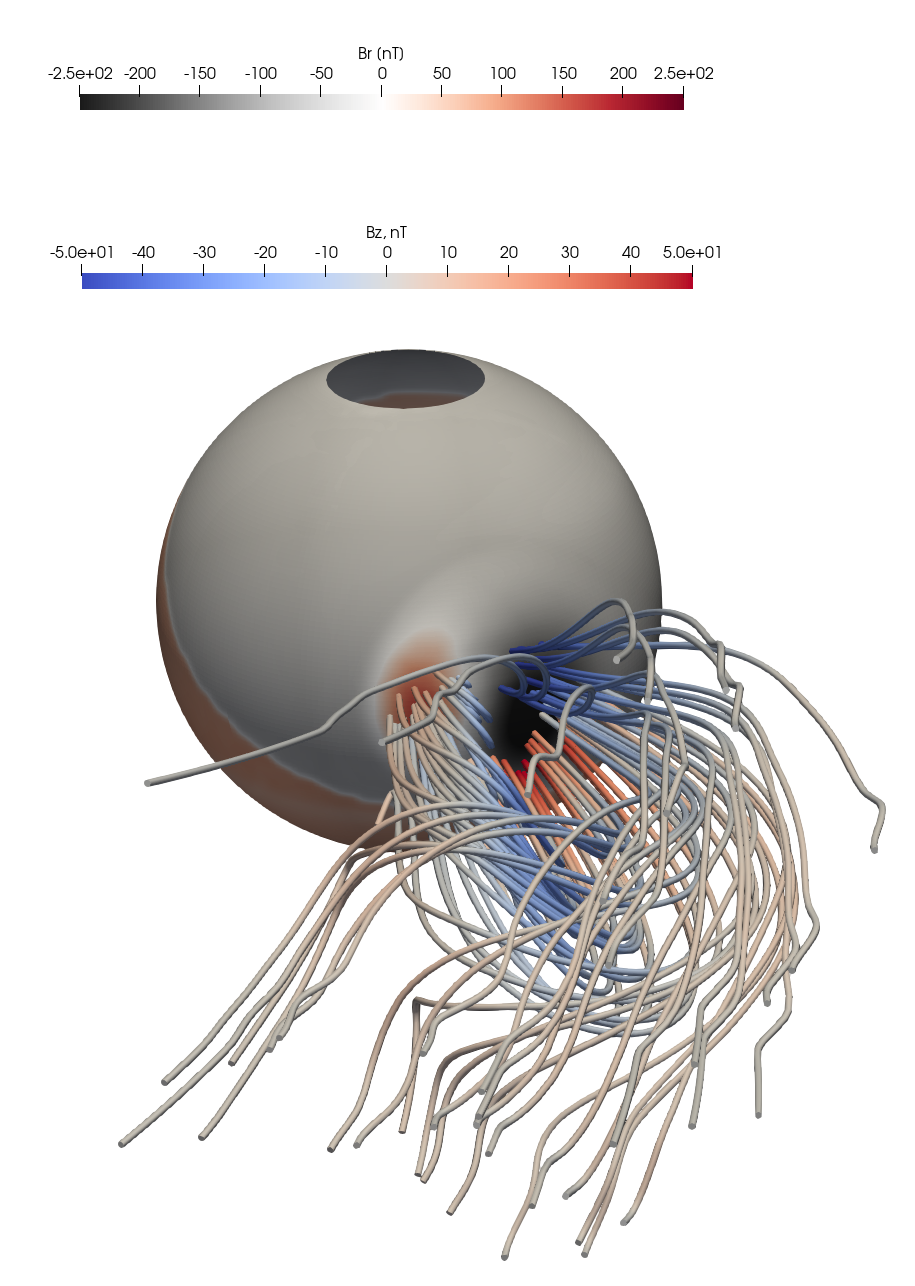}
\caption{CME magnetic field lines at t = 19 hr after the start of emergence. Field lines are traced from the emergence area and colored by $B_z$ (red-blue). The sphere at $21.5 \, R_S$ shows $B_r$ (red-grey) in a CME and background solar wind.}
\label{fig:3Demerge}
\end{figure}

To connect patterns seen in $B_r$, $B_{\theta}$, and $B_{\phi}$ in the CME cross-section in Figure \ref{fig:ibcgl} with the three-dimensional magnetic structure of the emerging GL CME shown in Figure \ref{fig:tethFR}, Figure \ref{fig:3Demerge} shows CME magnetic field lines traced from within the circle area after 19 hr since the emergence began. Because the CME magnetic field and interplanetary field are superimposed (equation \ref{eq:Bfield}), CME magnetic field lines, especially those at the CME periphery, instead of being closed as in the GL solution (Fig. \ref{fig:tethFR} C, F) become connected to the interplanetary field. The dipole-like pattern of $B_r$ at the inner boundary (top row in Fig. \ref{fig:ibcgl}) indicates predominantly radially directed magnetic field lines of a stretched flux rope as seen from the 3D structure in Fig. \ref{fig:3Demerge}. The $B_{\theta}$ pattern (second row in Fig. \ref{fig:bc}) represents winding magnetic field lines with the south-north component at CME periphery (negative $B_{\theta}$ regions at the sides)  and significant north-south component inside the CME (positive $B_{\theta}$). This is consistent with the winding of the magnetic field lines in the chosen GL magnetic topology shown in Fig. \ref{fig:tethFR} (C), (F).  The north-south directed field lines inside the CME is a region with the negative $B_z$ as shown in Figure \ref{fig:3Demerge} by the blue-shaded field lines inside the CME colored by the $B_z$ component.  
This is an extended region because the GL tethered flux rope stays attached to the GAMERA boundary as the CME emerges. The positive $B_{\phi}$ (third row in Fig. \ref{fig:ibcgl}) component reflects the East-West direction of the CME field lines due to the MFR orientation in the ecliptic plane. Overall, the magnetic structure of a CME propagating in the GAMERA model is consistent with the initial GL magnetic structure that we defined for this CME event.
%it takes 21 hours for the tethered flux rope to pass the inner boundary

%(since in ecliptic plane $B_{\theta}$ and $B_z$ differ by a sign, this region correspond to the negative $B_z$) 
The radial plasma velocity in the CME (second row from the bottom in Figure \ref{fig:ibcgl}) decreases with time until it reaches values in the solar wind background (provided by the WSA-ADAPT model) and is then kept at the background values.  This temporal behavior is expected since we add the CME and solar wind speeds in the emergence area, and the CME's speed decreases over time due to its self-similar expansion in the GL model. In the case shown here, the plasma velocity in a CME rapidly drops linearly from $800\, km/s$ to $550\, km/s$ within the first 3 hours of the emergence after which, it remains constant at the original background wind speed. Density enhancement within the CME remains constant throughout the emergence phase (bottom row in Figure \ref{fig:ibcgl}).

%CME evolution in the interplanetary medium reflects more the influence of large scale physical environment of the solar wind than self-similar expansion.

\begin{figure*}
\centering
\includegraphics[scale = 0.55]{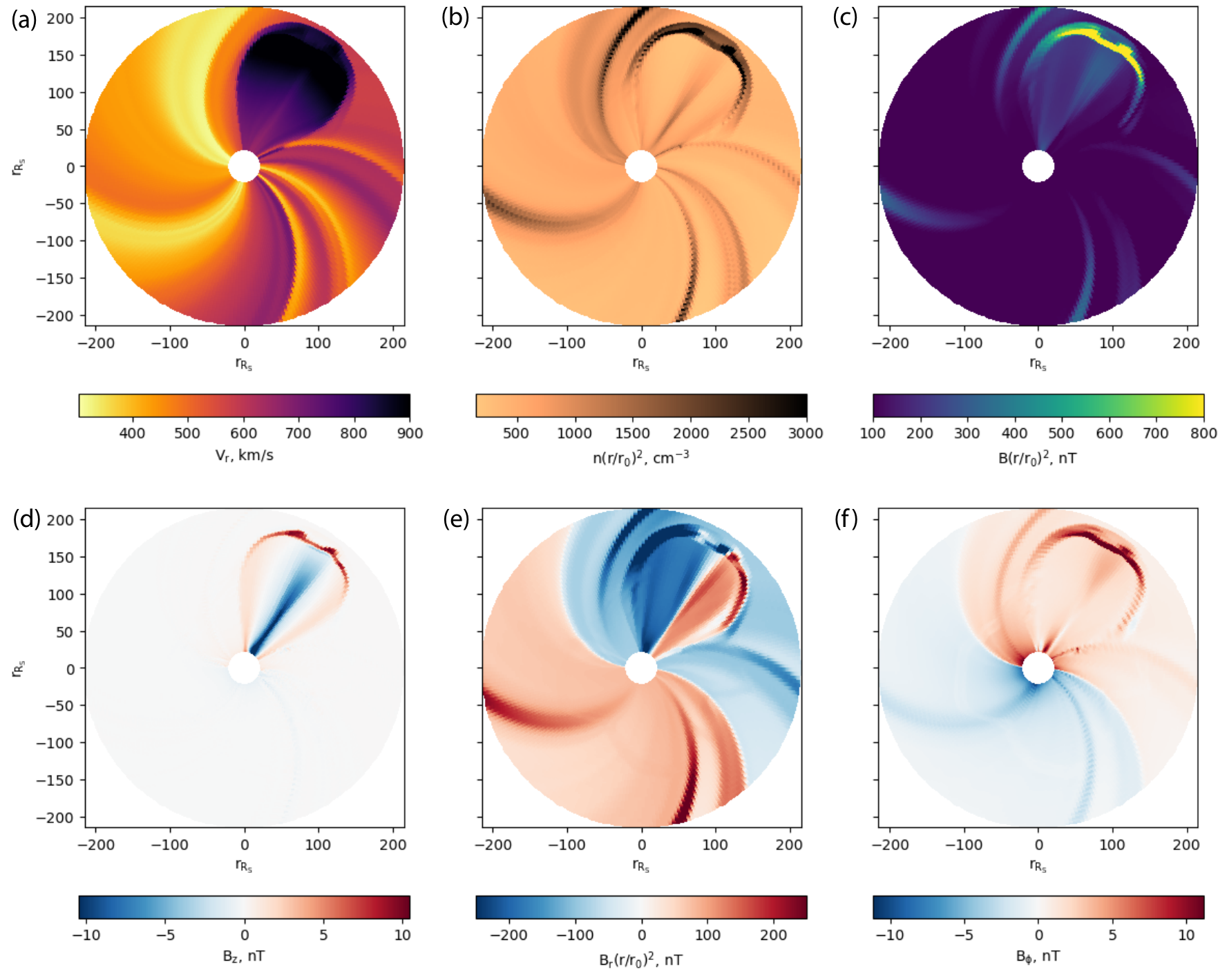}
\caption{(a) Radial velocity $V_r$, (b) scaled number density $n(r/r_0)^2$, (c) scaled magnetic field magnitude $B(r/r_0)^2$, and magnetic field components $B_z$ (d), $B_r(r/r_0)^2$ (e), $B_{\phi}$ (f) in the equatorial slice of a 3D simulation at t=36 hr after CME began its emergence. GL model parameters for the CME are given in Table \ref{tab:1} with the direction of propagation in the solar equatorial plane (Case 1).}
\label{fig:cmeecl}
\end{figure*}

Figure \ref{fig:cmeecl} shows a large-scale view of the simulated CME propagating in the solar wind as viewed in the solar equatorial plane (planar slice for $\theta=90^{\circ}$) at the $t=36\, hr$ after the CME flux rope has emerged through the inner boundary 
$R=21.5\, R_S$. At this time, the CME-driven shock approaches the outer boundary of the simulation domain at $215\, R_S$. 

As discussed above, since the CME flux rope emerges into a relatively uniform solar wind, the CME propagates without being strongly distorted by the interaction with the background solar wind. The overall GL magnetic structure is preserved as the CME propagates in the inner heliosphere, as can be seen from Figure \ref{fig:cmeecl}. 
In the GL solution for the tethered flux rope topology chosen for this CME event, the magnetic field lines remain attached to the origin (Sun) during the entire self-similar expansion (Fig. \ref{fig:tethFR}). Hence, the CME stays magnetically attached to the GAMERA inner boundary as it continues to emerge into the heliosphere. At the back side of the CME, magnetic field lines stretch predominantly in the radial direction, as shown by the $B_r$ plot (panel (e)). Meanwhile, magnetic field lines in the CME are twisted in such a way that they are directed south-north at the periphery of the CME (pointing out of the plane as indicated by the red color in panel (d)) and north-south inside the CME (pointing into the plane as indicated by the blue color in panel (d)). The tethered flux rope topology produces an extended region of negative $B_z$ in the CME, which is consistent with in-situ observations at Earth.
%The plasma speed gradually decreases from the CME front to the back, as expected, because in the GL CME model, the self-similar radial speed decays locally. The CME expands between the two stream interaction regions. In these regions, seen as spiral structures bounding the CME, the solar wind with intermediate speeds interacts with the slow solar wind streams, developing compression regions.

\begin{figure*}
\includegraphics[scale = 0.27]{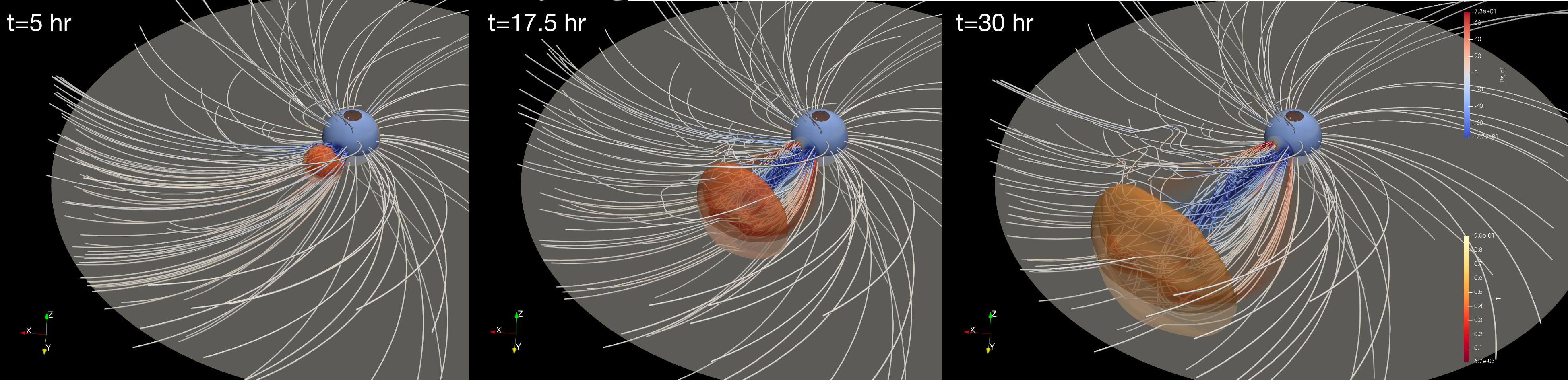}
\centering
\caption{Distorted structure of the heliospheric magnetic field as the CME with an internal flux rope with a large negative $B_z$ region propagates to 1 AU. Snapshots in time starting from the CME emergence at $21.5 \, R_S$ are shown.
Field lines traced from the emergence area at the inner boundary and an equatorial slice are color coded with the $B_z$ value. The GAMERA inner boundary sphere is color coded with the $B_r$ value. The temperature iso-surface $T=0.5$ MK ahead of the CME shows the CME sheath.  GL model parameters are given in Table 1, Case 1.}
\label{fig:cme3d}
\end{figure*}

\begin{figure*}
\centering
\includegraphics[scale = 0.29]{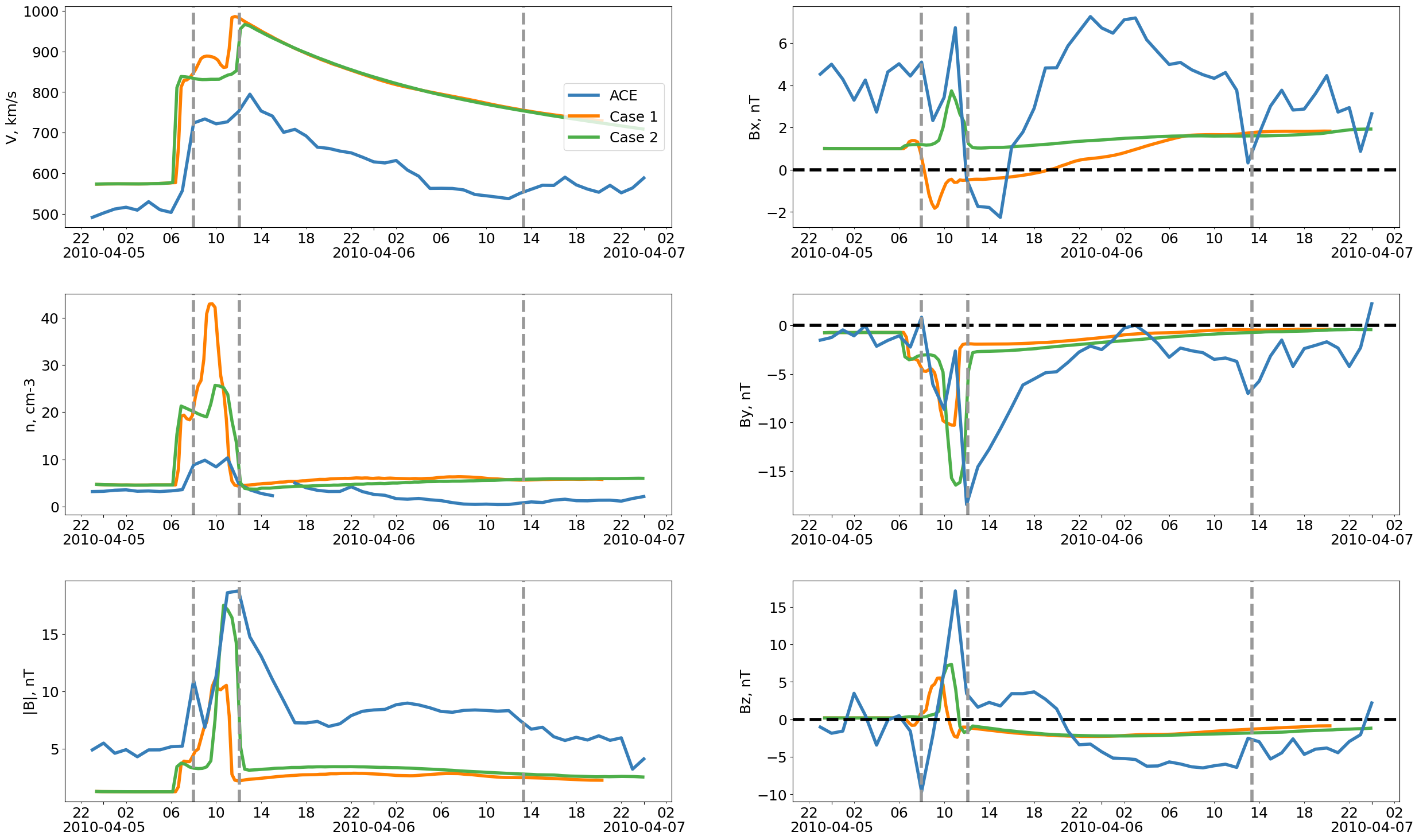}
\caption{ Time series of plasma speed, number density, magnetic field magnitude, and magnetic field components. {\it Blue} curves show hourly ACE in-situ observations, {\it Orange} curves show values extracted from GAMERA simulation at Earth in Case 1 with the ecliptic CME propagation, {\it Green} curves show values extracted from GAMERA simulation in Case 2 with the southern CME propagation. Vertical lines - from left to right - mark the CME shock arrival, the start and the end of the CME magnetic cloud.}
\label{fig:eclinsitu}
\end{figure*}

\begin{figure}
\includegraphics[scale = 0.53]{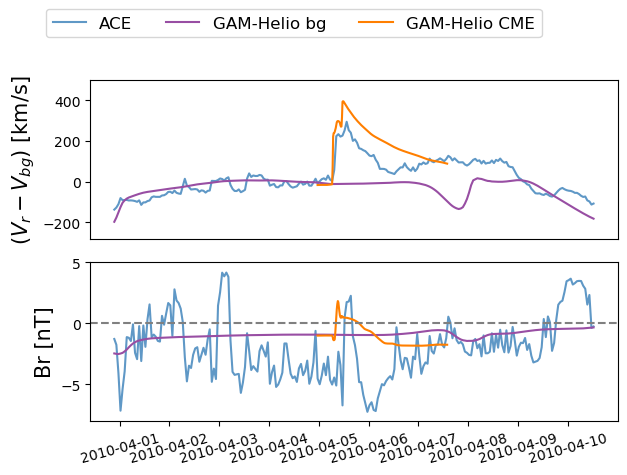}
\caption{Comparison of the ACE data (blue), GAMERA solar wind background simulation (purple) and GAMERA-GL simulation with the propagating CME (orange). Top: speed relative to the background solar wind speed just before the CME; bottom: radial component of the magnetic.}
\label{fig:compzoom}
\end{figure}

The CME flux rope creates a large-scale magnetic disturbance in the heliospheric magnetic field, producing closed magnetic field lines, increasing the length of the field lines, and connecting magnetically high and low latitudes. Figure \ref{fig:cme3d} shows snapshots of the magnetic field structure in the inner heliosphere for different times as the CME with the internal flux rope propagates and interacts with the interplanetary magnetic field. On the east side of the CME (as viewed from Earth), the closed magnetic field lines of the CME and the open magnetic field lines of the ambient interplanetary field are directed oppositely and cancel each other out. As a result,  some of the CME closed magnetic field lines become open into the interplanetary space, particularly those at the periphery of the CME. These magnetic field lines have an ``S''-shape similar to a scenario of interchange reconnection \citep{crooker2012interchange}.   A radially extended region with southward-directed magnetic field lines inside the CME is depicted by a blue shade in the equatorial plane. 

\subsection{CME parameters at Earth}\label{subsec:cmeatEarth}

Temporal variations of magnetic field components during Earth's passage of a CME depend on several factors including the initial CME magnetic structure leaving the Sun, its orientation, possible erosion en route to Earth, and a path along which Earth intersects the CME. 
We evaluate how well our models with the prescribed GL tethered flux rope for the 3 April, 2010 CME, the E-W orientation, and two propagation directions reproduce observed temporal variations of the magnetic field components. Figure \ref{fig:eclinsitu} shows plasma speed, density, magnetic field strength and components as functions of time as observed by the ACE spacecraft (blue curves) and extracted from the GAMERA simulations at Earth location for two directions of the CME propagation. Orange curves show GAMERA time series for the in-ecliptic CME propagation (shown by yellow arrow in Figure \ref{fig:steadystate} b). Green curves show GAMERA time series for the southward direction (shown by the white arrow in Figure \ref{fig:steadystate} b)). Gray vertical dashed lines -- from left to right -- show the CME shock arrival, the end of the CME sheath region, and the end of the CME magnetic ejecta in the ACE data, respectively. We mark these boundaries according to changes in observed plasma parameters and magnetic field at the shock, sheath, and ejecta boundaries determined in the analysis of in-situ measurements of this CME event in \citet{mostl2010stereo}.

In the model, the CME shock arrives $\sim$ 2 hours earlier than in the data. ACE observed the CME sheath lasting for $\sim$ 4 hours, while in the model, the sheath lasts $\sim$ 6 hours. The sheath is filled with dense, hot plasma and a compressed interplanetary magnetic field. Across the sheath, the modeled profiles show variations in plasma velocity, density, and magnetic field magnitude that are similar to the in-situ observations. Velocity increases slightly (case 1) or stays almost unchanged (case 2), density is enhanced, and the magnetic field increases, peaking at the boundary of the sheath. For both simulation cases, the model overestimates velocity and density values and underestimates magnetic field compression within the sheath compared to the observations.

The two GAMERA simulations with the same CME tethered flux rope and orientation but with two different CME propagation directions reveal distinct behaviors of the magnetic field components within the CME, as Earth samples different paths through the CME's magnetic structure.  
In the case of the in-ecliptic propagation (orange curves in Figure \ref{fig:eclinsitu}), the path goes through the middle of the CME. The modeled $B_x$ time profile shows a change from negative to positive values. The $B_y$ component remains negative throughout the CME's duration, while the $B_z$ component displays a prolonged period of negative values. In the case of the southward propagation direction (green curves in Figure \ref{fig:eclinsitu}), Earth is cutting through the northern edge of the CME.  The $B_y$ and $B_z$ components demonstrate behaviors similar to those seen in the in-ecliptic propagation. However, the $B_x$ component remains mostly positive in the ejecta lacking the rotation from negative to positive, as seen in observations and the simulation with in-ecliptic propagation. We note that we did not strive to match the magnitudes of magnetic field components that show large variations in the ejecta but rather capture their polarity, i.e., changes in their sign. 

To contrast the steady-state solar wind background simulation (shown in Figure \ref{fig:ACEcomp}) with the simulation including the CME, Figure \ref{fig:compzoom} compares the two solutions (steady-state in purple and CME solution in orange) with the in-situ ACE data (shown in blue) during the period 1-10 April around the CME passage. The top panel shows plasma speed with the speed of the solar wind background just before the CME subtracted. In contrast to the steady-state simulation, the GAMERA solution with the CME captures the sharp increase in the speed at the CME arrival and the gradual decrease during CME passage at Earth. The decrease in the model lasts longer, likely due to a slightly higher CME velocity and a more extended flux rope in the radial direction in the simulation. The bottom panel shows the radial magnetic field component. The GAMERA CME solution, shown in orange, shows the CME magnetic field disturbance with a shape similar to the in-situ data but weaker and shifted in time due to the CME in the model arriving earlier than in the data.
%
%
%
%
%
%DISCUSSION
%
%
%
%
%
%
%
%

\section{Discussion} \label{sec:Discussion}

The results presented in Section \ref{sec:results} show, that if we assume a southward direction of propagation, as observed in the COR imager near the Sun (Figure~\ref{fig:WL_obs}) and adopted in previous simulations with the hydrodynamic model for this CME \citep{wood2010}, the GAMERA-GL simulation shows a consistent behavior for only two out of the three magnetic field components between the observed 
 time series and extracted simulation data at Earth. Conversely, for the in-ecliptic propagation, the model shows variations of all three magnetic field components in the ejecta consistent with observations in terms of changing or preserving the sign.

It should not be surprising that the in-ecliptic simulation agrees better with the in-situ data. Figure~\ref{fig:WL_obs} clearly shows that the CME is expanding northward from about $7 R_S$ and has crossed the ecliptic plane by 20-30 $R_S$. The shape of the CME in the COR1-A and HI1-A fields of view is very different. This may be due to the CME rotation \citep[as reported by][]{Vourlidas_etal2011} or CME expansion along the ecliptic. The post-CME loop arcade has the appearance of an inverted 'J', which indicates the shape of the ejected MFR, at least to zeroth order. It seems quite plausible then that the expansion of the northern part of the MFR drives the northward expansion of the event as a whole. To the south, the CME expansion is restricted due to the presence of a southern polar coronal hole. The in-ecliptic CME propagation appears to be a reasonable assumption to check with the GAMERA-GL simulations that begin at $21.5 R_S$, presumably after the CME experienced the rotation or expansion in the corona. A better fit between the model and the data at Earth in this case, namely the changing sign of the $B_x$ component, a lasting negative $B_y$ component, and an extended period of a negative $B_z$ component in the CME, means that the GL magnetic flux rope with the given orientation and direction approximates well the magnetic structure of the 3 April, 2010 CME near 20 $R_S$.  

% large-scale magnetic structure of CME at 20 R_S is unknown, and unconstrained because of the lack of in-situ magnetic field observations for the CME near the Sun. To help with this gap, we utilized in-situ magnetic field observations at 1 AU, in particular changes of polarity of the three component and their duration, which helped us to choose unambiguously the appropriate magnetic topology among a range of topologies that the GL solution can produce. The tethered flux rope yields a long duration of the CME at Earth and magnetic field variations consistent with observations, particularly a long-lasting period of the negative $B_z$ component as seen in Figure .. . A spheromak or a half-spheromak is often assumed when modeling ICMES. In the case of the simple spheromak, the models the model would have yielded a significantly shorter duration of the event at Earth, and the $B_z$ component in the ejecta would have quickly flipped from negative values back to positive values.  making a spheromak CME model not appropriate for this event. 

The large-scale magnetic structure of CMEs at $\sim 20 R_S$ remains unconstrained due to the absence of routine in-situ magnetic field observations near the Sun. For the 3 April, 2010 CME, we have leveraged the in-situ magnetic field observations at 1 AU, focusing on changes in polarity of the three components and their durations in the ejecta. This approach has enabled us to unambiguously identify the appropriate magnetic topology from a range of possibilities that the GL solution can produce. The tethered flux rope model yields a prolonged duration of the CME at Earth and magnetic field variations consistent with observations, particularly an extended period of the negative $B_z$ component, as illustrated in Figure \ref{fig:eclinsitu}. Previous simulations of CMEs in the inner heliosphere considered spheromak or half-spheromak configurations. However, a spheromak CME would result in a significantly shorter duration of the event at Earth, with the $B_z$ component in the ejecta quickly transitioning from negative values back to positive values. Thus, a spheromak CME model is an inappropriate choice for this event. 

In general, representing a complex shape and magnetic structure of the CME in the outer corona with simplified analytical models -- such as a sphere or a tear-drop shape with internal twisted magnetic field lines -- is an extremely challenging problem for inner heliosphere models with CMEs starting from 0.1 AU. \citet{palmerio2023modeling} modeled a large event on 9 July, 2015 and demonstrated that different CME models may be suitable depending on the observer's position in the heliosphere or the specific CME characteristics of interest, such as the arrival time or magnetic field variations. 

When simulating the emergence of CMEs at 0.1 AU, it is important to take into account that plasma conditions within a CME may be significantly different from those in the ambient solar wind. 
Boundary conditions in inner heliosphere models, set on a sphere at 21.5 $R_S$ above the Alfv\'en surface in the solar wind, imply super-fast plasma flows, meaning that the solar wind speed is larger than the speed of the fast magnetosonic wave. While this assumption is justified for the solar wind flow, it may not hold true for flows inside the CME. For instance, this assumption breaks down for slow and strongly magnetized CMEs, potentially resulting in sub-fast flows at the inner boundary. In all simulations performed for the CME event studied in this work, the CME emergence occurred within the super-fast regime. However, to simulate CME emergence with internal sub-fast flows, an implementation of more general boundary conditions \citep{tarr2024simulating} would be necessary. 

%Depending on the magnitude of the radial stretching transformation applied to a spheromak, the GL model can produce a tear-drop-shaped region with some part of the original sphere contracted to the origin (center of the Sun). During the expansion in the corona, the tear-drop structure remains connected to the origin in a way that the trailing magnetic field lines are purely radial. When such a CME structure emerges through the boundary of the inner heliosphere model and propagates further to 1 AU, it stays magnetically connected  to the inner boundary of the simulation. At the boundary, the trailing radial field lines of the CME are surrounded by  radial open magnetic field lines of the interplanetary magnetic field. Such GL magnetic topologies are suitable for CME events that have magnetic field lines connected to the Sun and are typically associated with counter-streaming electrons \citep{gosling1987bidirectional}. For CMEs that lack such signatures, a spheromak or detached tethered spheromak topologies of the GL model would be more applicable. 

%\hl{EXPAND A BIT that in this study we superimpose fields}. An alternative approach is formulating boundary conditions for a CME emergence without adding the CME magnetic field and the pre-existing heliospheric magnetic field. It involves the construction of the potential solution for the heliospheric magnetic field that becomes tangential to the emerging CME structure. We will examine and implement such an approach in a separate study. 

%
%
%
%
%
%SUMMARY and CONCLUSIONS
%
%
%
%
%
%
%
%
\section{Summary and Conclusions} \label{sec:Summary}
In this work, we have presented the new coupling of the GAMERA inner heliosphere model with the Gibson-Low (GL) CME model to simulate the propagation of a CME with an internal magnetic field in the solar wind beyond the critical surface. The GL CME solutions include a range of magnetic topologies for a CME -- from a magnetic spheromak to a magnetic arcade -- providing flexibility in finding matching magnetic structures to observed CMEs.
The implementation of the model coupling works smoothly for an arbitrary solar wind map provided by the WSA-ADAPT model for the solar wind background simulation and a wide range of input CME parameters such as angular width, speed, maximum magnetic field strength, and magnetic topology.
With a spatial resolution sufficient for resolving large-scale structures of the CME flux rope, CME sheath, and shock, as well as large-scale streams in the solar wind background, such simulations require modest computational resources compared to full MHD Sun-to-Earth CME simulations. 

We used the newly developed model to simulate the geoeffective CME event that occurred on 3 April, 2010. This CME was an isolated event in the solar corona (without interaction with other CMEs) and well-observed with coronagraphic and heliospheric images from SOHO/LASCO and both STEREO/SECCHI imaging suites. These images helped us to constrain the input parameters for the GL CME emerging into the inner heliosphere model at $21.5 R_S$, particularly the CME angular width, speed, direction of propagation, and orientation of the CME flux rope. To determine the GL magnetic topology of the CME, we examined the time profiles of the magnetic field components measured in situ by the Wind spacecraft during the CME passage. Observations show the rotation of the magnetic field in the CME and a prolonged region with a negative $B_z$ component, indicating the tethered flux rope topology in the GL CME model. Coronal and heliospheric images show two different CME propagation directions, so we performed simulations for two directions - in-ecliptic and southward.

Our results can be summarized as follows:
\begin{enumerate}
\setlength{\itemsep}{3pt}
  \item 
  We find that the GL tethered flux rope is an appropriate magnetic topology for representing the 3 April, 2010 CME event as it approaches $\sim 20\, R_S$  in the outer solar corona. This topology produces time variations of the magnetic field components during Earth`s passage of the CME and the CME duration that are consistent with the in-situ ACE measurements.
  
  \item 
  In the case of in-ecliptic CME propagation, the simulated variations of the three magnetic field components during the  Earth`s passage of the CME are consistent with in-situ observations in terms of the duration of the southward $B_z$ component and changes in the sign of the components. This direction of the CME propagation is supported by the heliospheric HI1 imaging observations.
  
  \item With the CME moving southward, supported by COR images near the Sun, two of three magnetic field components, $B_y$ and $B_z$, agree with the \textit{ACE} in-situ data. However, the $B_x$ component consistently shows positive values throughout the CME passage, in contrast to observations that demonstrate a variation from negative to positive.
  
  \item 
  Our simulations unambiguously point to the $0^{\circ}$ orientation of the CME flux rope in the outer solar corona (meaning that the axis of symmetry of an initial spheromak before the stretching is oriented in the N-S direction), or, in other words, the flux rope has an East-West orientation. While \citet{wood2011} showed that the unique orientation for this CME could not be inferred from the STEREO and SOHO/LASCO observations, they concluded that the East-West orientation fits better for the heliospheric propagation of the CME observed in HI1 images. Our results support the East-West orientation, demonstrating the importance of such simulations in determining CME flux rope orientations.
\end{enumerate}

\acknowledgments
This work was supported by the NASA LWS Science Program award number 80NSSC17K0685.
E.P was also supported by the AFOSR YIP program award number FA9550-21-1-0276. V.G.M. was supported by the AFOSR grant FA9550-21-1-0457. A.V. was supported by NASA grant 80NSSC22K0970. E.W. was supported by 80NSSC21K1745. We acknowledge high-performance computing resources of the NSF-sponsored Derecho: HPE Cray EX System (\url{https://doi.org/10.5065/qx9a-pg09}) and Cheyenne: HPE/SGI ICE XA System provided by NCAR's Computational and Information Systems Laboratory (CISL). All in-situ data used in this study are available at NASA's CDAWeb data repository (\url{https://cdaweb.gsfc.nasa.gov/}). The simulation data are of large volume and are therefore available upon request.

\bibliography{CMEpaper}

\end{document}